\documentclass[prb,12pt,superscriptaddress]{revtex4-1}
\usepackage{pdfpages}
\usepackage{amsmath}    
\usepackage{graphicx}   
\usepackage{fancyhdr}
\usepackage{datetime}

\begin{document}

\title{Switchable friction enabled by nanoscale self-assembly on graphene}

\author{Patrick Gallagher}
\author{Menyoung Lee}
\affiliation{Department of Physics, Stanford University, Stanford, CA 94305, USA}
\author{Francois Amet}
\affiliation{Department of Physics, Duke University, Durham, NC 27708, USA}
\affiliation{Department of Physics and Astronomy, Appalachian State University, Boone, NC 28608, USA}
\author{Petro Maksymovych}
\author{Jun Wang}
\affiliation{Center for Nanophase Materials Sciences, Oak Ridge National Laboratory, Oak Ridge, TN 37831, USA}
\author{Shuopei Wang} 
\author{Xiaobo Lu}
\author{Guangyu Zhang}
\affiliation{Institute of Physics, Chinese Academy of Sciences, Beijing, 100190, China}
\author{Kenji Watanabe}
\author{Takashi Taniguchi}
\affiliation{National Institute for Materials Science, 1-1 Namiki, Tsukuba, 305-0044, Japan}
\author{David Goldhaber-Gordon}
\email{goldhaber-gordon@stanford.edu}
\affiliation{Department of Physics, Stanford University, Stanford, CA 94305, USA}

\maketitle

\textbf{Graphene monolayers are known to display domains of anisotropic friction with twofold symmetry and anisotropy exceeding 200 percent\cite{Choi2011}. This anisotropy has been thought to originate from periodic nanoscale ripples in the graphene sheet\cite{Choi2011,Choi2012,Choi2014}, which enhance puckering around a sliding asperity\cite{Lee2010} to a degree determined by the sliding direction. Here we demonstrate that these frictional domains derive not from structural features in the graphene, but from self-assembly of atmospheric adsorbates into a highly regular superlattice of stripes with period 4 to 6 nm. The stripes and resulting frictional domains appear on monolayer and multilayer graphene on a variety of substrates, as well as on exfoliated flakes of hexagonal boron nitride. We show that the stripe-superlattices can be reproducibly and reversibly manipulated with submicron precision using a scanning probe microscope, allowing us to create arbitrary arrangements of frictional domains within a single flake. Our results suggest a revised understanding of the anisotropic friction observed in graphene and bulk graphite\cite{Rastei2013,Rastei2014} in terms of atmospheric adsorbates.}

Nanometer-scale surface textures with long-range order often give rise to pronounced frictional anisotropy. These textures sometimes originate from crystal structures: periodic tetrahedral reversals in the antigorite lattice create nanoscale surface corrugations, which generate the anisotropic friction that governs certain seismic processes\cite{Campione2013}. A large frictional anisotropy similarly arises for some quasicrystal intermetallics, whose surfaces are textured by atomic columns\cite{Park2005}. Rotationally aligned adsorbates, including molecules in organic films\cite{Last1996} and tilted alkyl chains in lipid monolayers\cite{Liley1998}, also form ordered nanotextures with associated anisotropic friction. Rotational symmetry of the host surface permits multiple stable molecular orientations, yielding frictional domains with anisotropy along different axes. 

From a technological standpoint, nanometer-scale systems with such multistability are appealing platforms for switches or memories. Bistable states in redox centers\cite{Gittins2000}, rotaxane molecules\cite{Cavallini2003}, and iron clusters\cite{Loth2012} can be addressed and switched using scanned probes, enabling dense information storage. Multistable nanotextures could find application in nanoelectromechanical systems if the friction-producing textures could be dynamically controlled, as in biomimetic tapes with magnetically actuated micropillars\cite{Northen2008}. Existing schemes for tuning friction at submicron scales include Fermi level modulation in silicon\cite{Park2006} and mechanical oscillation of a sliding contact\cite{Socoliuc2006}---nonhysteretic techniques which require maintenance of a voltage or oscillation, a disadvantage for circuitry.

In this work, we identify friction-producing nanotextures that naturally form on graphene exposed to air, and exploit their multistability to hysteretically switch friction with submicron precision. Using high-resolution atomic force microscopy (AFM), we directly image superlattices of nanoscale stripes that produce the reported\cite{Choi2011} anisotropic friction on exfoliated graphene. These nanotextures strongly resemble patterns of nitrogen adsorbates observed at the interface between water and graphite\cite{Lu2012}, and we induce apparently identical nanotextures on flakes of hexagonal boron nitride by thermal cycling. Consistent with the adsorbate picture, we can rapidly and predictably reorient the frictional domains by scanning a probe tip along the flake in a chosen direction---a departure from nanoassembly techniques\cite{Tseng2011} like dip-pen nanolithography\cite{Piner1999} and nanografting\cite{Xu1997}, for which writing a different ``color" requires submerging the sample in a different ``ink." 
  
To image friction, we measure the deflection (diving board motion) and torsion (axial twist) of a scanned AFM cantilever in light contact with the sample. The deflection signal primarily contains topographic information, while the meaning of the torsion signal depends on scan direction. For lateral scanning (motion perpendicular to cantilever axis; Fig. 1b, lower panel), the torsion measures lateral tip-sample forces commonly interpreted as friction forces. In this ``friction imaging" mode, tip-sample forces transverse to the scan direction result in deflection, contributing spurious topographic signals. When the cantilever is scanned longitudinally (Fig. 1c, lower panel), the torsion signal directly measures tip-sample forces transverse to the scan direction. For an isotropic surface, this ``transverse force" signal is zero. 
  
As reported previously\cite{Choi2011,Choi2012}, the friction signal of exfoliated monolayer graphene flakes on silicon dioxide reveals up to three distinct domains of friction despite a featureless topography signal (Fig. 1a,b). The domains vary in size from tens of nanometers to tens of microns, and produce sharp contrast in transverse force, confirming their anisotropic character (Fig. 1c). Tapping mode AFM images taken with ultrasharp tips within the different domains (Fig. 1d) reveal periodic stripes along axes rotationally separated by 60$^\circ$ (angular orientation does not measurably vary within a given domain; see Supplementary Fig. 1). To within experimental error (typically $\pm$0.2 nm), stripe period (typically $\sim$4 nm) does not change across a sample, although we have observed global changes in stripe period after thermal cycling (e.g., from 4 to 6 nm in Supplementary Fig. 2). Peak-to-trough stripe amplitude ranges between 10 and 100 pm, but strongly depends on tip conditions and oscillation parameters.

The observed frictional anisotropy of a given domain respects the symmetry of the stripe-superlattice. The friction signal approximately tracks the cosine of the angle between scan axis and stripes (Fig. 1e)---friction is \emph{maximized} when the two are aligned---while the transverse force is zero when the stripes are perpendicular or parallel to the scan axis, as required by symmetry (Fig. 1f). In between these zeros, the transverse force changes sign so as to guide the sliding tip toward the low friction axis (lower panel, Fig. 1c). We conclude that the stripes in graphene produce the observed friction anisotropy, similarly to friction-producing nanotextures in other systems\cite{Campione2013,Park2005,Last1996,Liley1998}.

The stripes are not unique to monolayer graphene on SiO$_2$. We observe stripe domains and anisotropic friction on graphene flakes up to 50 nm thick (the maximum thickness investigated) without change in stripe period and with minor change in magnitude of frictional anisotropy (Supplementary Fig. 3), as well as on graphene flakes on different substrates (Supplementary Fig. 4). Stripe domains can also form on exfoliated flakes of hexagonal boron nitride (hBN) on SiO$_2$: single crystals show at most three distinct domains of anisotropic friction (Fig. 2b,c), each characterized by a different orientation of stripes, whose typical period is $\sim$4 nm (Fig. 2d). As for graphene, the friction signal is maximized when scanning along the stripes. But whereas we observe stripes on nearly all graphene flakes as exfoliated, to form stripes on hBN typically requires a cryogenic thermal cycle (such as immersion in liquid nitrogen; see Methods). Our variable-temperature AFM study found stripes to form on hBN upon cooling from 300 K to 250 K, although vacuum conditions likely influence the evolution with temperature (Supplementary Fig. 5).

The behavior of stripes on epitaxial heterostructures of graphene and hBN implies that stripes on both materials share a common origin. The nearly perfect rotational alignment between stacked lattices\cite{Yang2013} results in a moir\'e pattern with lattice constant $\sim$14 nm in regions where graphene has grown on the hBN (Fig. 3a). Despite this additional superstructure, stripes form on exposed layers of both graphene and hBN with no measurable difference in period, and often appear to maintain phase across a graphene/hBN boundary. Furthermore, using the moir\'e pattern to infer lattice orientation\cite{Tang2013} (Fig. 3b), we find that the stripes run along the armchair axes of both crystals in all 25 epitaxial heterostructures and 5 mechanically assembled heterostructures that we studied (Supplementary Fig. 6). 

Previous studies have ascribed the anisotropic friction in monolayer graphene to periodic ripples in the graphene sheet induced by stress from the substrate\cite{Choi2011,Choi2012,Choi2014}. While our data confirm the presence of periodic structure, the extreme similarity of the stripes on graphene and hBN---materials with different bending stiffness and response to stress\cite{Singh2013}---suggests that the stripes are adsorbates rather than features of the crystals themselves. The orientation of the stripes further rules out periodic ripples, which would produce a high friction axis perpendicular to the stripes\cite{Choi2011,Choi2012}, and a zigzag stripe axis\cite{Choi2014,Ma2011}---both opposite to our findings. 

On the other hand, surfactant molecules are known to self-assemble into stripes along the armchair axes of graphite\cite{Manne1995}; molecular length and Debye screening\cite{Wanless1996} determine the stripe period (4 to 7 nm), while anisotropic van der Waals interactions lead to crystallographic alignment. Atmospheric species produce similar stripes: crystallographically aligned stripes of 4 nm period were observed on graphite submerged in water, and correlated with the presence of dissolved nitrogen gas\cite{Lu2012}. Stripes of similar period were later observed in ambient on multilayer epitaxial graphene, and were attributed to nitrogen adsorbates stabilized by a water layer\cite{Wastl2013}. Gas enrichment at the interface between water and a hydrophobic surface is theoretically expected\cite{Dammer2006}, although no explanation has been given for the formation of stripes instead of a homogeneous layer. To this end, we note that a periodic arrangement of nitrogen-rich and water-rich regions is a generically anticipated consequence\cite{Seul1995} of the competition between domain wall energy, which favors large domains, and dipolar repulsion between water molecules, which favors small domains, assuming that the water dipoles align when near the graphite surface.

We propose that the stripes on graphene and hBN are self-assembled atmospheric adsorbates, likely nitrogen and water. This hypothesis explains the appearance of armchair stripes on graphene and hBN, both hydrophobic surfaces (when hBN is sufficiently flat\cite{Li2012}) with hexagonal symmetry. It also explains the absence of stripes in ambient scanning tunneling microscopy (STM) with atomic resolution (Supplementary Fig. 7): the adsorbates are disturbed by the pressure of the STM tip\cite{Magonov2008}. We suggest nitrogenic rather than organic stripes since our pristine surfaces (cleaved or annealed; see Methods) likely lack the concentration of organics required for self-assembly\cite{Manne1995}.

Recent work resolved nanoscale stripes in the transverse force response of bulk graphite, and ascribed them to a novel puckering-induced stick-slip friction process\cite{Rastei2013}. These stripes produced domains of anisotropic friction\cite{Rastei2014} like those on graphene and hBN. We suggest a reinterpretation of these data in terms of atmospheric adsorbates, which would unify our understanding of anisotropic friction in graphite, graphene, and hBN.

Adsorbates can sometimes be mechanically manipulated by AFM\cite{Tseng2011}, raising the possibility of patterning friction on these materials. For monolayer graphene on SiO$_2$, scanning at the low normal force used for imaging (1 nN) often minimally affects the frictional domains, but scanning at high normal force (30 nN) reproducibly reorients the domains (Fig. 4a). We devised two standard approaches for domain manipulation (Fig. 4b). The ``brush stroke" consists of raster scanning a rectangular window at high normal force; we retract the tip after every line so that it only scans the sample in one direction. Brush strokes produce reproducible results---often a domain flop---that depend on the scan angle and the initial ``canvas" domain. For scan angles near the canvas stripe axis, the canvas switches to the domain with stripes next closest to the scan axis (Fig. 4c). Our second approach is to ``erase" the canvas domain within a rectangular scan window by rapid, back-and-forth scanning at high normal force. This mode destabilizes the domains within the scan window, leaving only the most stable domain, determined primarily by local strain and partly by scan axis. Although strain gradually varies across the flake (see discussion below), erasing still produces deterministic results within a specific region.

The brush stroke and eraser allow us to rapidly create patterns of friction with submicron precision. Without optimizing our procedure, creating a block letter `S' 5 microns tall using the eraser took 16 minutes, while creating a `U' using brush strokes took 36 minutes (Fig. 4d and Supplementary Movie 1). After writing, the pattern gradually decayed: here the `S' widened, while the `U' narrowed (Fig. 4e). We wrote the same pattern in different parts of the flake, and found that whether a domain grew or shrank with time, and how rapidly it evolved, depended on position. The absence of other obvious symmetry-breaking mechanisms suggests that local strain induced by the substrate determines the relative stability of the domains. Domain stability in turn determines the effective resolution of our patterning technique: although we can write crisp lines 100 nm wide in some parts of a flake, in other parts these features only persist for minutes before decaying to match the canvas domain. Additionally, while we can pattern friction on several different monolayer graphene flakes, others show only weak response to both patterning modes described; the strain field in these flakes likely strongly favors the local canvas domain, making it difficult to switch.

Whether patterning friction is possible on thicker crystals requires further investigation. Our first attempts indicate that domains can be rewritten with the eraser or brush stroke, although the resulting domains are not as sharp as on monolayer graphene. Proximity to the substrate could be stabilizing the stripes, allowing for more flexible control of domain shape. Our work underscores the major role played by atmospheric adsorbates, rather than structural deformation, in determining friction on graphene and hBN---and perhaps on other hydrophobic surfaces, such as transition metal dichalcogenides\cite{Gaur2014}. The periodic perturbation from the adsorbates might open gaps at the superlattice energy or modify the Fermi velocity in graphene\cite{Park2008}, with measurable consequences for electronic properties of ultraclean graphene/hBN heterostructures\cite{Dean2010}.

\section*{Methods}

\subsection*{Sample preparation}
Flakes of graphene and hBN were prepared by mechanical exfoliation (3M Scotch 600 Transparent Tape or 3M Scotch 810 Magic Tape) under ambient conditions (40-60\% relative humidity) on silicon wafers with 90 nm or 300 nm of thermal oxide. The substrates were not exposed to any chemical processing following thermal oxidation. For graphene exfoliation, we used bulk crystals of both Kish graphite (Sedgetech, USA) and highly-oriented pyrolitic graphite (HOPG ZYA, SPI Supplies, USA), and observed no relevant differences in superlattice phenomena between these two graphite sources. For hBN exfoliation, we used bulk crystals provided by Kenji Watanabe and Takashi Taniguchi. We also prepared graphene flakes on other substrates (Supplementary Fig. 4), including SU-8 epoxy (MicroChem, USA), 200 nm of Au(111) on mica (Phasis, Switzerland), and 5 nm of Pt (electron-beam evaporation) on magnesium oxide (MTI, USA). 

We prepared epitaxial graphene heterostructures on oxidized silicon substrates by mechanical exfoliation of hBN followed by graphene growth at 500$^\circ$C by a remote plasma-enhanced chemical vapor deposition process described previously\cite{Yang2013}. We also mechanically assembled heterostructures of graphene on hBN using both wet\cite{Amet2013} and dry\cite{Wang2013} transfer methods. Polymer residues from the assembly process were removed by annealing samples in a tube furnace for 4 hours at 500$^\circ$C under continuous flow of oxygen (50 sccm) and argon (500 sccm); before removal to air, we allowed the samples to cool (5 to 10$^\circ$C min$^{-1}$) to below 100$^\circ$C under the same flow of oxygen and argon.

\subsection*{Thermal cycling}
We found stripes to appear on our samples after thermal cycling to liquid nitrogen temperatures or below using a variety of methods. Most commonly, and specifically for the sample shown in Fig. 2, we immersed the sample in liquid nitrogen for one to five minutes and then removed it to atmosphere, and blew off the condensation with dry air. This procedure would almost always produce stripes on graphene, hBN, or graphene/hBN heterostructures. In other cases, we loaded the sample in the vacuum chamber of a cryostat---either a cryogen-free dilution refrigerator or a Quantum Design PPMS---and thermal cycled to a base temperature between 25 mK and 100 K. Cooling and warming rates varied between 1 and 30 K min$^{-1}$. We warmed up the samples under various atmospheres including moderate vacuum, helium gas, or nitrogen gas; in all of these cases (over ten different samples cycled in the dilution refrigerator or PPMS) we found stripes on every flake or heterostructure (totaling several tens) that we checked.

The epitaxial heterostructure in Fig. 3 was not cycled to low temperature: the sample had stripes after removal from the growth furnace. Some of our assembled heterostructures (Supplementary Fig. 6) required a low temperature thermal cycle to produce stripes after the oxygen/argon anneal, although in other cases we observed stripes without cryogenic treatment.  

\subsection*{AFM and STM measurements}
All images shown in Figs. 1-4 were taken with a Park XE-100 AFM under ambient conditions (40-60\% relative humidity) except for Figs. 4d and 4e, which were taken in 10\% relative humidity by flooding the chamber of the XE-100 with dry air. (We observed no significant difference in domain mutability or evolution between 10\% and 50\% relative humidity.) To resolve the stripes in tapping mode, we used sharp silicon probes (MikroMasch Hi'Res-C15/Cr-Au) with a nominal tip radius of 1 nm, a typical resonant frequency of 265 kHz, and a typical cantilever $Q$ of 400. For measurements in contact mode, we used silicon probes (MikroMasch HQ:NSC19/Al BS-15) with a nominal tip radius of 8 nm and a typical resonant frequency of 65 kHz. We used a normal force setpoint of 1 nN for all friction and transverse force imaging scans shown, with scan rates $\sim$10 $\mu$m s$^{-1}$. For domain manipulation, we used a normal force setpoint of 30 nN; for brush strokes, we used scan rates $\sim$30 $\mu$m s$^{-1}$, while for erasing, we used scan rates $\sim$300 $\mu$m s$^{-1}$. 

When imaging friction or transverse force, we collected torsion data for both forward-moving and backward-moving scans. To eliminate offsets in the friction and transverse force signals for Figs. 1e and 1f and Supplementary Fig. 3d, we subtracted backward images from forward images and divided by two. All friction or transverse force images shown are just the forward scan, with any torsion offset eliminated by subtracting the average of forward and backward torsion values on SiO$_2$.

To study stripe formation with changing temperature (Supplementary Fig. 5), we used an Omicron varible-temperature AFM/STM operating in ultrahigh vacuum (UHV; $8 \times 10^{-11}$ mbar). Samples were not baked in UHV prior to experiments. The sample stage was cooled by a copper braid attached to a cold sink held at low temperature by continuous flow of liquid nitrogen; by this method we achieved a base temperature of 110 K. We used the same sharp probes as for ambient AFM (MikroMasch Hi'Res-C15/Cr-Au). In ultrahigh vacuum, the cantilever $Q$ reached 5000, which significantly restricted scan speed for tapping mode; we therefore used on-resonance frequency-modulation mode, imaging at a typical frequency shift of -30 Hz. For all images, we applied a DC tip-sample bias to nullify the contact potential difference.

STM measurements (Supplementary Fig. 7) were carried out under ambient conditions using the Park XE-100. We prepared our tip by mechanically cutting a Pt/Ir wire and scanning the sample at high bias voltages until we achieved atomic resolution of the graphene lattice. 

\subsection*{Error bars and lateral calibration}
All values quoted for moir\'e period and angular orientation are extracted from the FFT of the AFM images. AFM images of all heterostructures described in this study are corrected for thermal drift by performing an affine transformation to produce regular moir\'e hexagons (we used the free software Gwyddion, available at gwyddion.net). All error bars reflect the full width at half maximum (FWHM) of the peaks in the FFT; for instance, 12.0 $\pm$ 0.5 nm means that the FWHM of the peak maps to 1 nm in real space. The lateral scale of the Park XE-100 was calibrated by measuring the moir\'e period of graphene/hBN heterostructures grown by van der Waals epitaxy, in which the graphene and hBN lattices are nearly perfectly aligned, and defining this period (averaged over several samples) to be 13.6 nm. This definition corresponds to the assumption made in Supplementary Fig. 6 that the lattice constants for hBN and graphene are $a_{\rm hBN} =$ 0.25 nm and $a_{\rm graphene} = a_{\rm hBN}/$1.018. The lateral scale of the Omicron variable-temperature AFM was calibrated to the lateral scale of the Park XE-100 by measuring the moir\'e pattern of the same sample in both systems.

\section*{Acknowledgments}
We gratefully acknowledge Byong-man Kim and Ryan Yoo of Park Systems for verifying the presence of stripes in our samples using their Park NX-10 AFM. We thank Daniel Wastl for carefully reading our manuscript, and for encouraging us to re-examine whether the stripes we observed were caused by periodic structural ripples or self-assembled adsorbates. We thank Trevor Petach and Arthur Barnard for other helpful discussions. Sample fabrication and ambient AFM/STM were performed at the Stanford Nano Shared Facilities with support from the Air Force Office of Science Research, Award No. FA9550-12-1-02520. Variable-temperature AFM studies were conducted at the Center for Nanophase Materials Sciences, which is a DOE Office of Science User Facility; our use of the facility was supported by the Center for Probing the Nanoscale, an NSF NSEC, under grant PHY-0830228. S.W., X.L., and G.Z. acknowledge support from the National Basic Research Program of China (Program 973) under grant 2013CB934500, the National Natural Science Foundation of China under grants 61325021 and 91223204, and the Strategic Priority Research Program (B) of the Chinese Academy of Sciences under grant XDB07010100. K.W. and T.T. acknowledge support from the Elemental Strategy Initiative conducted by the MEXT (Japan). T.T. acknowledges support from JSPS Grant-in-Aid for Scientific Research under grants 262480621 and 25106006.

\section*{Author contributions}
P.G. identified the stripes, performed all experiments, and wrote the paper. P.G. and F.A. fabricated the assembled heterostructures. M.L., F.A., and D.G.-G. discussed data and experimental directions, and assisted in writing the paper. P.M. and J.W. supported the variable-temperature AFM measurements. S.W., X.L., and G.Z. grew the epitaxial graphene/hBN heterostructures. K.W. and T.T. grew the bulk hBN crystals.

\section*{Competing financial interests}

The authors declare no competing financial interests.

\clearpage

\begin{figure}
\centering
\includegraphics[width=6.2in]{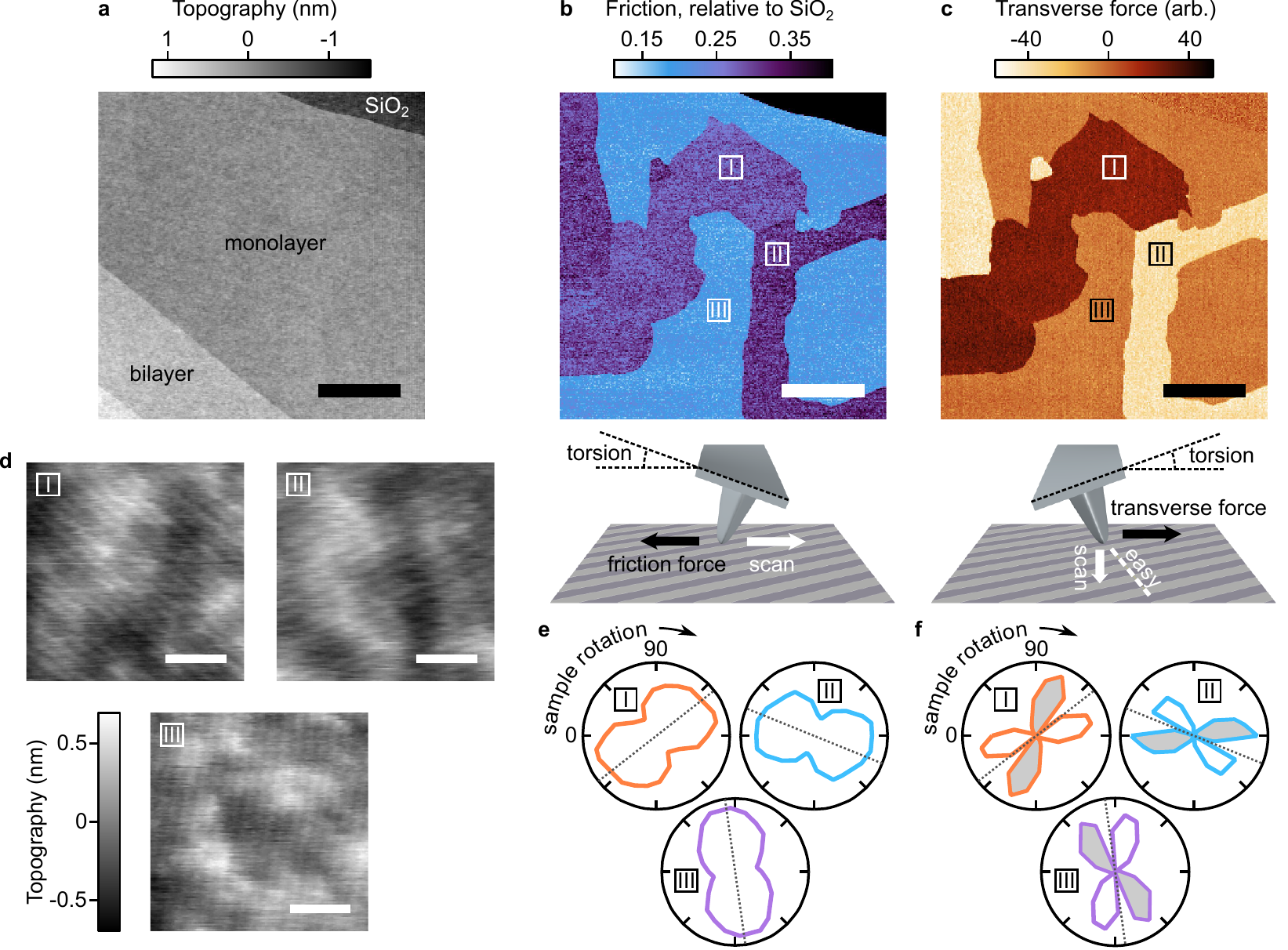}
\caption{\linespread{1}\selectfont{}{\bf Stripes on exfoliated graphene.} {\bf a}, Contact mode topography scan of a graphene flake on silicon oxide, showing monolayer, bilayer, and trilayer regions. Scale bar: 3 $\mu$m. {\bf b}, Simultaneously recorded friction signal (upper panel), showing three distinct domains of friction labeled I, II, and III. Lower panel: cartoon of the friction imaging mode. The cantilever is scanned laterally, and friction between the tip and sample produces the measured torsion of the cantilever. {\bf c}, Transverse force signal (upper panel) from the same region as in {\bf b}, measured by recording the torsion while scanning the cantilever longitudinally (lower panel). Surface anisotropy pushes the tip toward the local ``easy" axis, creating a transverse force that twists the cantilever. {\bf d}, Tapping mode topography scans of the graphene monolayer, taken within each of the three domains. Each domain is characterized by stripes of period 4.3 $\pm$ 0.2 nm along one of three distinct axes rotationally separated by 60$^\circ$. Scale bars: 20 nm. {\bf e}, Friction relative to SiO$_2$ for each domain as a function of clockwise sample rotation angle; zero degrees corresponds to the orientation shown in {\bf a}-{\bf c}. For each polar plot, the origin and circumference correspond to relative friction values of 0.15 and 0.4, respectively. Dotted lines indicate the sample rotations at which the stripes shown in {\bf d} are parallel to the scan axis. The friction signal is approximately sinusoidal, with the highest friction produced when stripes are parallel to the scan axis. {\bf f}, Transverse force signal for each domain as a function of clockwise sample rotation angle. Unshaded and gray-shaded regions indicate positive and negative transverse signals, respectively. The origin of each polar plot is zero. The transverse signal for a given domain switches sign as the stripe axis rotates through the lateral axis. 
}
\end{figure}

\begin{figure}
\centering
\includegraphics[width=6.2in]{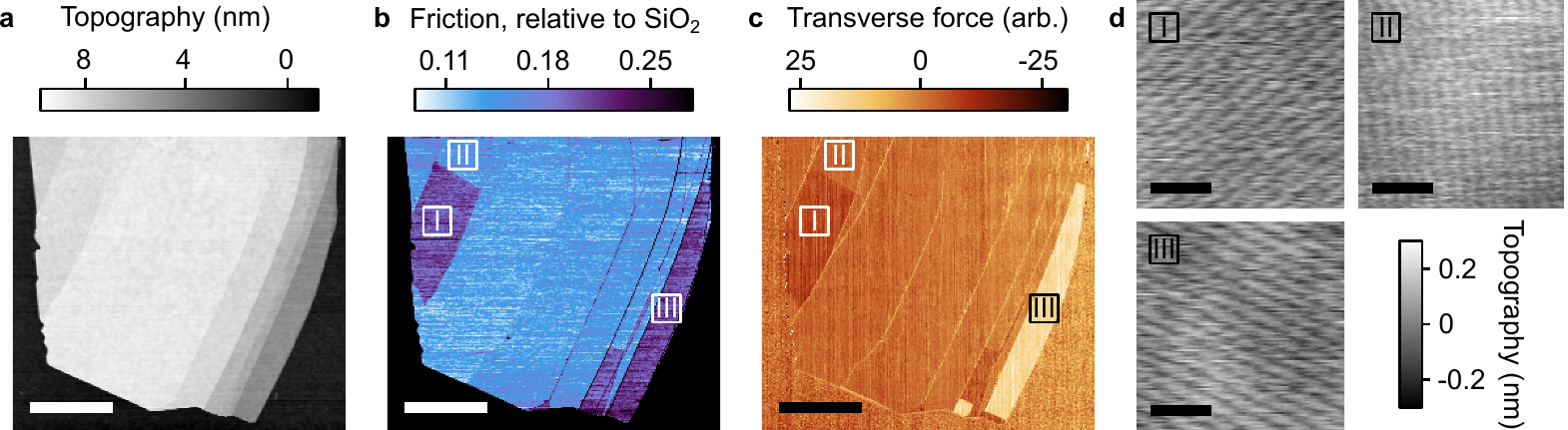}
\caption{\linespread{1}\selectfont{}{\bf Stripes on exfoliated hBN.} {\bf a}, Contact mode topography scan of a terraced hBN flake, thickness 5 to 9 nm, after thermal cycling in liquid nitrogen. Scale bar: 5 $\mu$m. {\bf b} and {\bf c}, Simultaneously recorded friction signal ({\bf b}) and separately recorded transverse force signal ({\bf c}) showing the presence of three distinct domains (I, II, and III). The contrast between I and III is weak in friction, but strong in transverse force. {\bf d}, Tapping mode topography scans of the three domains, taken in the regions indicated in {\bf b} and {\bf c}. Each domain is characterized by stripes of period 4.7 $\pm$ 0.2 nm along one of three distinct axes rotationally separated by 60$^\circ$. Scale bars: 20 nm.
}
\end{figure}

\begin{figure}
\centering
\includegraphics[width=3in]{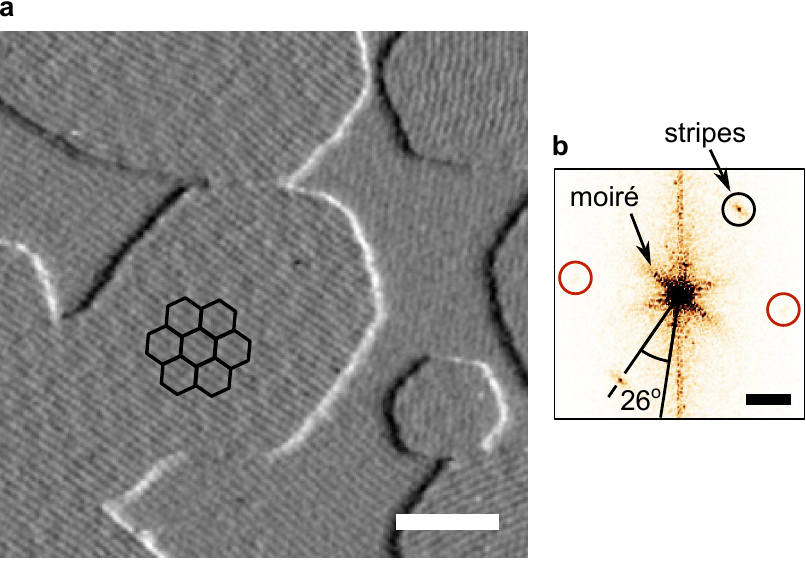}
\caption{\linespread{1}\selectfont{}{\bf Orientation of stripes on graphene and hBN.} {\bf a}, Tapping mode topography image of graphene islands grown by van der Waals epitaxy on exfoliated hBN. The image has been differentiated along the horizontal axis for clarity. Graphene islands can be distinguished from the hBN surface by the presence of a moir\'e pattern, which is partially outlined in black for one of the grains. The sample surface is covered with stripes of period 4.3 $\pm$ 0.1 nm, oriented along one of three distinct axes rotationally separated by 60$^\circ$. The stripe period is the same on graphene and hBN, and the stripes frequently appear to cross the graphene/hBN boundary without a phase slip. Scale bar: 50 nm. {\bf b}, Fast Fourier transform (FFT) of the topography signal used to produce {\bf a}. The moir\'e pattern within the graphene grains appears as a sixfold-symmetric pattern with segments extending $\sim$70 $\mu$m$^{-1}$ from the origin; these protruding segments are parallel to the momentum-space moir\'e lattice vectors. The dominant stripe domain on graphene and hBN produces a pair of isolated points in the FFT, one of which is circled in black. The stripe axis is rotated 26 $\pm$ 4$^\circ$ from the moir\'e lattice vectors, indicating that the stripe axes are nearly aligned with the armchair axes of the graphene and hBN. The quoted angular precision reflects the width of the moir\'e peaks; we also expect a few-degree systematic error in the angular estimate, since a misalignment between graphene and hBN lattices of 0.1$^\circ$---a reasonable expectation for van der Waals-epitaxial heterostructures\cite{Tang2013}---would rotate the moir\'e pattern by 4$^\circ$ with respect to the graphene lattice. The small area of nearly vertical stripes in {\bf a} produces a pair of points, circled in red, which can barely be seen with this colorscale. Scale bar: 100 $\mu$m$^{-1}$.
}
\end{figure}

\begin{figure}
\centering
\includegraphics[width=6.2in]{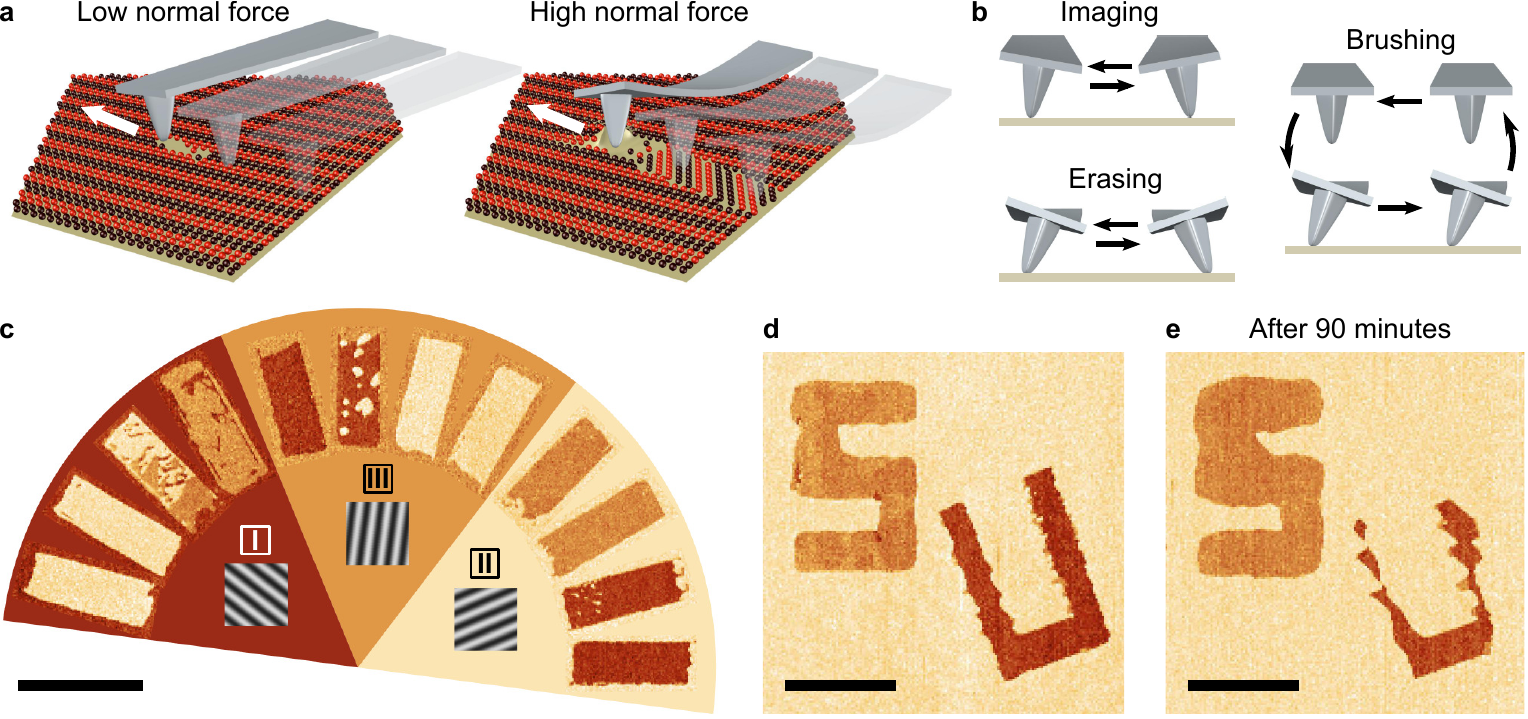}
\caption{\linespread{1}\selectfont{}{\bf Rewritable friction on monolayer graphene.} {\bf a}, Cartoon illustrating the response of the striped adsorbates to the scanning tip. At low normal force, the tip minimally disturbs the stripes as it scans the surface, and the stripe structure rapidly heals. At high normal force, the stripe structure is heavily disturbed, creating a new stripe domain in the wake of the scanning tip. {\bf b}, Summary of our scanning modes. For imaging, we rapidly scan the cantilever back and forth at low normal force while slowly moving it in the direction perpendicular to the fast scan axis. The erasing mode is identical, but at high normal force. For a brush stroke, we raster-scan the cantilever such that the tip only moves in one direction when in contact with the sample. After scanning each line, we lift the cantilever, move it to the start of the next line, and touch down again. {\bf c}, Domain switching as a function of scan angle on the monolayer flake studied in Fig. 1, rotated as in Fig. 1a-c. The image shown is a collage of twelve transverse force images, each taken after executing a single 3 $\mu$m by 1 $\mu$m brush stroke on a canvas composed initially of a single domain. For each canvas domain we show four brush strokes nearly parallel with the canvas stripes, where each brush stroke is directed radially outward from the origin of the semicircle. The brush strokes steer the canvas domain toward the domain whose stripes are next nearest the brush axis. Scale bar: 3 $\mu$m. {\bf d}, Transverse force image immediately after writing block letters `S' and `U' in domains III and I, respectively, on a canvas of domain II (same flake and orientation as in a). The block letter `S' was written by ``erasing," while the `U' was written with a brush stroke. Scale bar: 3 $\mu$m. {\bf e}, Transverse force image of the same area, taken 90 minutes later. The `S' (domain III) has expanded into the canvas, while the `U' (domain I) has decayed. 
}
\end{figure}

\clearpage


\section*{Supplementary material}

# Supplementary Information for: Switchable friction enabled by nanoscale self-assembly on graphene

Patrick Gallagher,[1] Menyoung Lee,[1] Francois Amet,[2,3] Petro Maksymovych,[4]

Jun Wang,[4] Shuopei Wang,[5] Xiaobo Lu,[5] Guangyu Zhang,[5] Kenji

Watanabe,[6] Takashi Taniguchi,[6] and David Goldhaber-Gordon[1, *]

[1]*Department of Physics, Stanford University, Stanford, CA 94305, USA*

[2]*Department of Physics, Duke University, Durham, NC 27708, USA*

[3]*Department of Physics and Astronomy,*

*Appalachian State University, Boone, NC 28608, USA*

[4]*Center for Nanophase Materials Sciences,*

*Oak Ridge National Laboratory, Oak Ridge, TN 37831, USA*

[5]*Institute of Physics, Chinese Academy of Sciences, Beijing, 100190, China*

[6]*National Institute for Materials Science,*

*1-1 Namiki, Tsukuba, 305-0044, Japan*

## UNIFORMITY OF STRIPE-SUPERLATTICE WITHIN A SINGLE DOMAIN

Single domains of stripes may span many microns. Here we illustrate the extreme uniformity of the stripe-superlattice by analyzing a representative large area scan (scan window approximately 600 nm) within a single domain on hBN (Supplementary Fig. 1a); results for graphene are similar. We have deliberately chosen an image that contains a phase slip to illustrate that small superlattice defects are possible—but we emphasize that phase slips are uncommon away from domain boundaries or structural defects, and that phase slips could be induced by the moving tip rather than being a feature of an unperturbed sample. The perceptible waviness of stripes on the graphene grain in the upper-left of Fig. 3a is another example of a defect. We only observe such waviness in epitaxial samples, and only near boundaries between graphene and hBN regions.

In Supplementary Fig. 1a, the peak-to-trough topographic amplitude of the stripes is $30 \pm 10$ pm throughout the scan window (Supplementary Fig. 1b). Angular resolution of the stripe axis is limited by slowly-varying scanner drifts, which manifest as smearing of the superlattice peak in the fast Fourier transform (FFT) along the direction perpendicular to the scan axis (Supplementary Fig. 1c). Here this smearing permitted an angular resolution of $4°$ (Supplementary Fig. 1d); while scan parameters can be optimized for slightly improved angular resolution (for instance, by increasing the scan speed to mitigate thermal drifts), dramatic improvement is difficult. To this few-degree precision, we observe no changes in stripe axis within a domain. The period over a large area can be measured with very high *precision*: a radial cut through the superlattice peak yields a period $6.43 \pm 0.04$ nm (Supplementary Fig. 1e). However, extremely slowly varing thermal drifts will cause an affine distortion of the image, which strongly impacts the *accuracy* of the measured period (and stripe axis). The large-area scan in Supplementary Fig. 1a required a particularly low scan rate, and is therefore particularly strongly affected by drift; smaller, faster scans of the same area reveal a stripe period of $4.6 \pm 0.2$ nm.

## VARIABILITY IN GLOBAL STRIPE PERIOD

We typically observe a global stripe period between 4 and 5 nm for stripes on both graphene and hBN, but we sometimes observe the period to change significantly after ther-

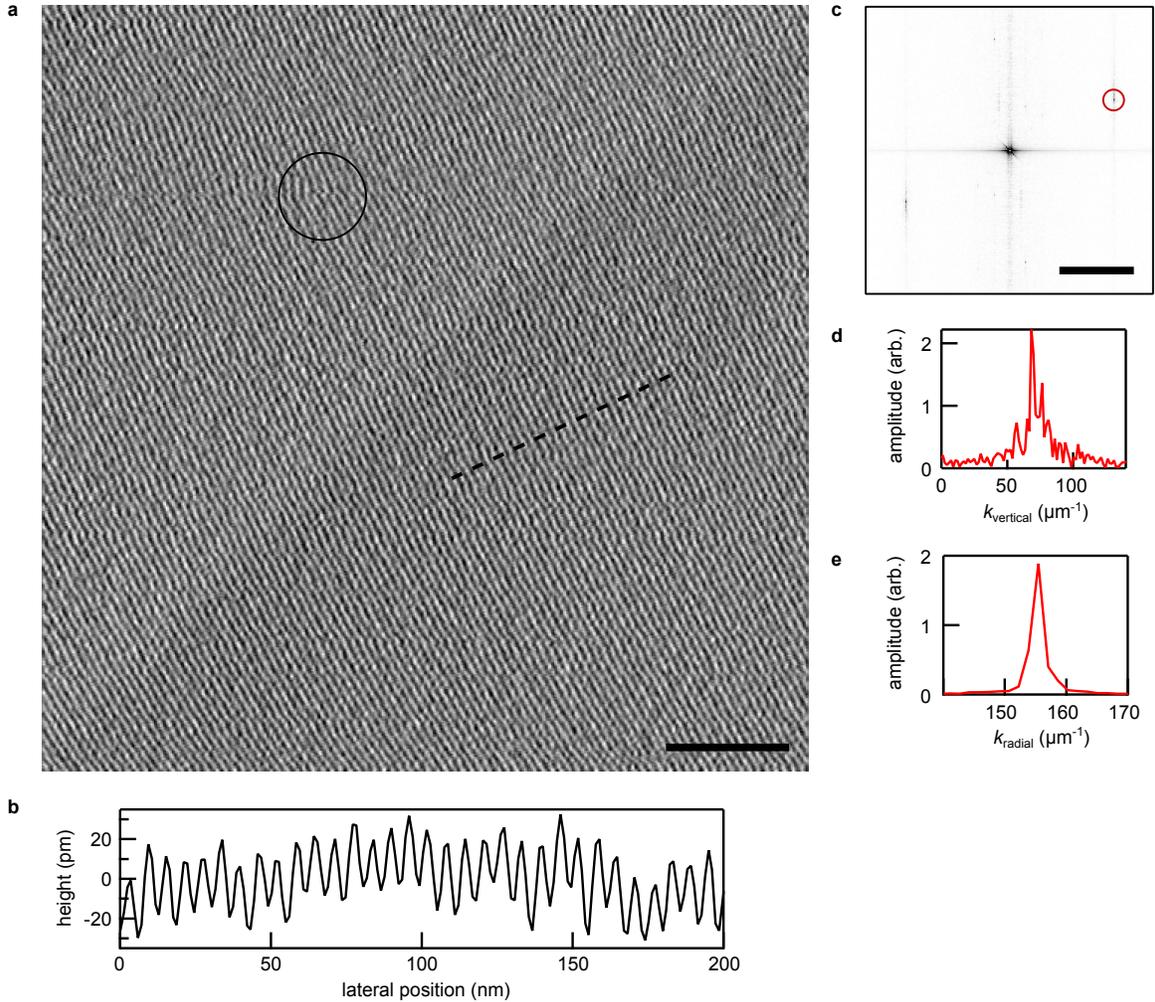

**Supplementary Figure** 1. Uniformity of stripe-superlattice within a single domain on hBN. **a**, Tapping-mode topography signal, differentiated along the scan axis (horizontal) to suppress the gentle topographic background (the broad feature running bottom left to top right is a smooth depression ∼100 pm deep). Stripe axis runs upper left to lower right. Black circle is centered around a phase slip: three parallel stripes merge into two. Scale bar: 100 nm. **b**, Height (undifferentiated) along black dashed line in **a**. Each point is averaged over 16 nm transverse to the black dashed line. **c**, FFT of the topography signal used to produce **a**. One of the two superlattice peaks is circled in red. Scanner drift smears the peaks along the vertical axis. Scale bar: 100 $\mu$m$^{-1}$. **d**, Vertical cut through the superlattice peak in **c**; $k_{\text{vertical}} = 0$ corresponds to the vertical center of panel **c**. FWHM is 10 $\mu$m$^{-1}$, which corresponds to a 4° error in estimation of the stripe axis. **e**, Radial cut through the superlattice peak in **c**; $k_{\text{radial}} = 0$ corresponds to the center of panel **c**. The stripe period is 6.43 ± 0.04 nm, where the error represents the FWHM of the peak (2 $\mu$m$^{-1}$).

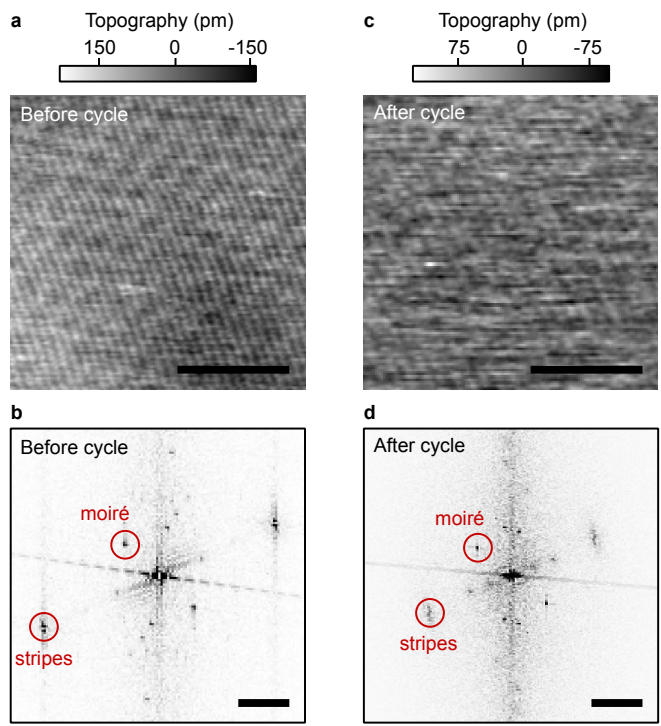

**Supplementary Figure** 2. Change in global stripe period after thermal cycling a sample. **a**, Topography signal and **b**, fast Fourier transform (FFT) of a mechanically assembled graphene/hBN heterostructure. Stripes of period $4.3 \pm 0.1$ nm are superimposed on a moiré pattern of lattice constant $11.9 \pm 0.4$ nm. **c**, Topography signal and **d**, FFT of the same sample after several thermal cycles between 10 K and 390 K in a cryostat (PPMS). The moiré lattice constant has not changed, as expected, but the stripe period is now $5.9 \pm 0.3$ nm throughout the sample. Topography scale bars: 50 nm. FFT scale bars: 100 $\mu m^{-1}$.

mal cycling a sample. For example, imaging one graphene/hBN heterostructure before (Supplementary Fig. 2a,b) and after (Supplementary Fig. 2c,d) thermal cycling between 10 K and 390 K in our PPMS revealed an increase in global stripe period of $1.6 \pm 0.3$ nm (here the measured change in period is very accurate, as we have used a moiré pattern in the sample to calibrate away thermal drifts). Within the adsorbate picture, these changes in period can be explained by the addition or removal of electrolytic impurities, which modify the Debye screening length and therefore the period; for instance, changing the salt concentration was shown[1] to modify the period of self-assembled surfactants on graphite by up to 2 nm.

## ANISOTROPIC FRICTION ON MULTILAYER GRAPHENE

Stripes appear on both monolayer and multilayer graphene flakes, and produce domains of anisotropic friction that extend across step-edges with minor change in frictional contrast (Supplementary Fig. 3a-c). Previous work reported the friction on monolayer graphene to be twice as large as the friction on "bulk-like" flakes (4 layers or thicker), an effect ascribed to increased puckering around the tip for few-layer graphene[2]. By comparison, frictional contrast in our images (Supplementary Fig. 3b,c) is dominated by the frictional domains, and we find minimal contrast between monolayer and bulk-like flakes within the same domain.

As we increase the normal force setpoint beyond the 1 nN normal force setpoint that we typically use for imaging, contrast between monolayer and bulk-like flakes grows (Supplementary Fig. 3d), approaching agreement with the values quoted in Ref. 2. These results are consistent with the presence of at least two distinct friction-producing effects: drag from an adsorbate layer (the stripes), which dominates at low normal force and produces friction anisotropy, and puckering of the flake, which more sharply increases with applied load and contributes isotropic friction. Ref. 2 evidently did not observe frictional domains, despite imaging at a 1 nN normal force setpoint with probe tips nominally identical to ours (although with a different cantilever stiffness). We suspect that some aspect of the sample preparation in Ref. 2 prevented formation of the stripe-superlattice.

## STRIPES ON GRAPHENE ON SUBSTRATES OTHER THAN SILICON OXIDE

To determine the relevance of the specific choice of substrate to the appearance of stripes, we exfoliated graphene flakes onto a variety of substrates. For graphene on 200 nm of Au(111) on mica, we observed stripes (Supplementary Fig. 4) on two flakes out of approximately ten that we checked. After thermal cycling by immersion in liquid nitrogen, we found stripes on the same two flakes, and still none on other flakes. On hardbaked SU-8 photoresist spun 15 microns thick, we found stripes on all (approximately ten) exfoliated graphene flakes that we checked. On platinum thin films (5 nm) evaporated on magnesium oxide, we found no stripes on any of approximately ten graphene flakes examined. In all cases, the stripes identified had the same approximate period and amplitude as those found

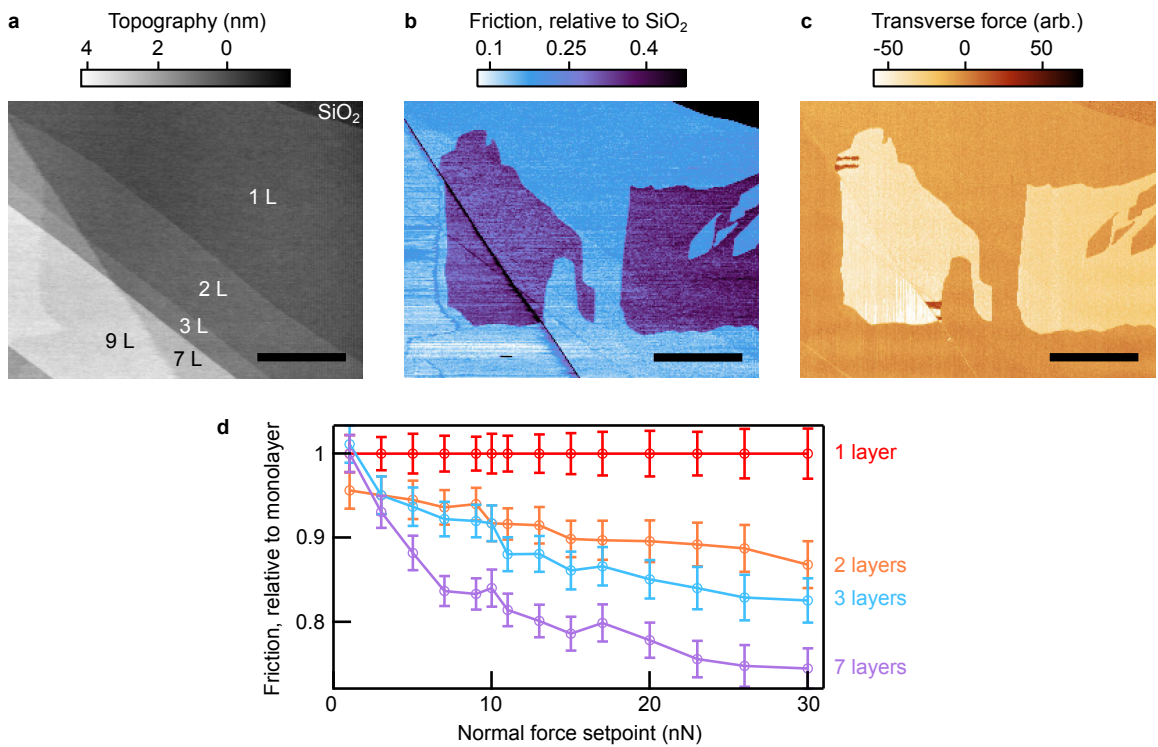

**a** Topography (nm)
**b** Friction, relative to SiO₂
**c** Transverse force (arb.)

SiO₂
1 L
2 L
3 L
9 L
7 L

**d**
Friction, relative to monolayer

1 layer
2 layers
3 layers
7 layers

Normal force setpoint (nN)

**Supplementary Figure** 3. Anisotropic friction on multilayer graphene flakes. **a**, Contact mode topography signal from a different region of the terraced flake shown in Fig. 1, with the sample oriented as in Fig. 1a-c. Text indicates the number of graphene layers in various regions of the flake. Normal force setpoint was 1 nN. Scale bar: 5 $\mu$m. **b**, Simultaneously recorded friction signal, showing primarily two domains which extend over several steps in the terrace. Friction contrast between the domains does not strongly change with flake thickness. **c**, Transverse force scan of the same region, also with normal force setpoint 1 nN. Again the contrast between domains is similar for regions of different thickness. **d**, Friction signal versus normal force setpoint for regions of the flake shown in **a-c** with 1, 2, 3, and 7 layers of graphene, all within the low friction domain. The data are normalized to the friction on the monolayer. At low normal force, the contrast between monolayer and 7 layers is minimal, but at higher normal force the contrast grows. Error bars reflect the standard deviation of the friction signal within each domain.

on graphene on SiO₂ substrates.

In accordance with our interpretation of the stripes as nitrogen adsorbates stabilized by a water layer, we hypothesized that substrate hydrophilicity might impact stripe formation on a given flake. To investigate this possibility, we measured the contact angle of water

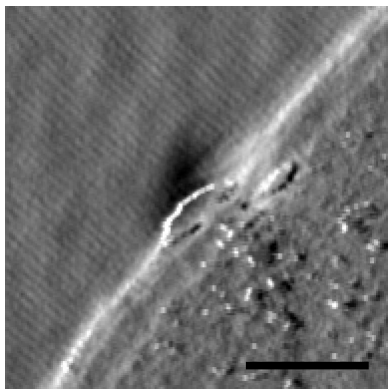

**Supplementary Figure** 4. Stripe-superlattice on a gold/mica substrate. Tapping-mode topography signal, differentiated along the scan axis (horizontal), at the edge of an as-deposited, few-layer graphene flake on gold on mica. Stripes are visible on the graphene (upper left half of image) with period 3.8 ± 0.2 nm; peak-to-trough amplitude in the topography signal is 20 pm. Scale bar: 50 nm.

droplets (5 $\mu$L; contact area much larger than flake size) on the various substrates described above, as well as on several pieces of $SiO_2$, and found that the contact angle was by itself not a good predictor of the presence of stripes on exfoliated graphene. For instance, both a platinum/magnesium oxide sample (no stripes on graphene) and an SU-8 sample (stripes on graphene) had a water contact angle of 80°, while the water contact angle on different $SiO_2$ samples with stripes ranged between 35° and 70°. The precise effect that the substrate and other environmental factors have on the self-assembly of stripes merits further investigation.

## VARIABLE TEMPERATURE AFM MEASUREMENTS

Using a variable temperature AFM (VT-AFM) operating in ultrahigh vacuum (UHV; see Methods), we scanned for stripes on an as-exfoliated hBN flake on $SiO_2$ while holding the temperature at successively lower setpoints starting from 300 K (Supplementary Fig. 5). The sample was not baked after loading in the UHV chamber, leaving a layer of ambient adsorbates on the flake. At 300 K, we could not resolve stripes on the sample, consistent with our typical results for as-exfoliated hBN scanned under ambient conditions. Once the sample had cooled to the first low temperature setpoint (250 K), we could resolve stripes, which persisted to 225 K, but could not be resolved again at 200 K or 175 K. We stepped the

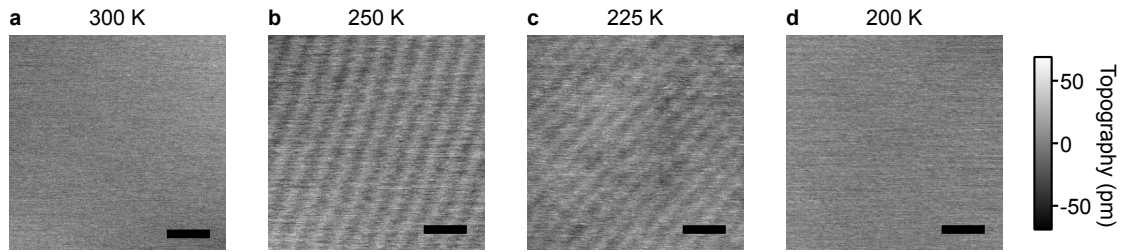

**Supplementary Figure** 5. Formation and suppression of stripes with decreasing temperature. **a–d**, Non-contact AFM topography of an hBN crystal 60 nm thick, as deposited on SiO$_2$, at various temperature setpoints as the temperature was lowered. Scale bars: 50 nm. At 300 K (**a**), no stripes are visible. At 250 K (**b**), stripes have formed, with peak-to-trough amplitude 15 pm. The stripes are still visible at 225 K (**c**); changes in the apparent period and stripe axis between 250 K and 225 K could result from thermal scanner drift, or from slightly differing scan locations. At 200 K (**d**), stripes can no longer be resolved. Horizontal features faintly visible in all scans, but especially in **a** and **d**, are artifacts of the horizontal scan axis.

temperature back up to 300 K and still could not resolve stripes at any of several temperature setpoints. We note that the images in Supplementary Fig. 5 are not corrected for thermal drift, which significantly affects the apparent angle and period of the stripes. Uncontrolled lateral movement ($\sim$1 $\mu$m) of the tip relative to the sample between temperature setpoints also prevented us from scanning precisely the same location at every temperature on the relatively featureless hBN flakes that we studied. For these reasons, we hesitate to make any claims about the apparent change of period and stripe axis between 250 K and 225 K in Supplementary Fig. 5.

We performed similar VT-AFM experiments on two other hBN flakes (on two separate oxidized silicon pieces) which had been cycled to low temperature in a separate cryostat before loading in the VT-AFM (again not baked after loading in the UHV chamber). On both samples, we could resolve stripes at 300 K in UHV prior to decreasing the temperature. In one sample, we directly lowered the temperature from 300 K to 110 K, and found no stripes at low temperature. In the other sample, we lowered the temperature to 250 K and still observed stripes, but found them to disappear upon lowering to 200 K. Stripes did not reappear in either sample at any temperature setpoint while warming up to 300 K. Furthermore, we did not observe stripes in ambient AFM scans of either sample after

removal from the VT-AFM.

While a detailed analysis of the temperature dependence of the stripes is beyond the scope of our work, we consistently found that in UHV, the stripes on hBN disappeared at and below 200 K and did not reappear upon warming up to room temperature. In contrast, thermal cycles in environments with higher pressure almost always produced stripes at room temperature (Methods). We suggest that the atmosphere surrounding the sample strongly impacts stripe formation.

## CRYSTALLOGRAPHIC ORIENTATION OF STRIPES

For heterostructures with close alignment between graphene and hBN (including some assembled heterostructures, and all epitaxial heterostructures), we extract the crystallographic orientation of the stripe axes using the calculated angular misalignment[3] between the moiré lattice vectors and the graphene lattice vectors. Our data rule out the possibility that the stripe axes are parallel to the graphene lattice vectors (zigzag axes), since the measured angle between the stripe axes and the nearest moiré lattice vector does not follow the expected trend as a function of moiré superlattice period (Supplementary Fig. 6a). In contrast, our angular misalignment data conform to expectations if we assume that the stripe axes are always armchair (Supplementary Fig. 6b).

## ABSENCE OF STRIPES IN STM

To help distinguish between explanations for the stripes, we studied epitaxial graphene on hBN under ambient conditions by both AFM and STM, with the STM tunneling parameters optimized for resolution of the atomic lattice and moiré pattern. The tapping mode AFM scans reveal stripes of topographic amplitude comparable to the moiré corrugation, but the STM scans in both constant current and constant height modes only reveal a moiré pattern with no stripes (Supplementary Fig. 7). These data are consistent with self-assembled stripes of adsorbates: the stripes would be disturbed by the STM tip, which sits Angstroms from the graphene surface when imaging the graphene lattice. If the stripes were structural ripples, we would expect to see them in both AFM and STM topography.

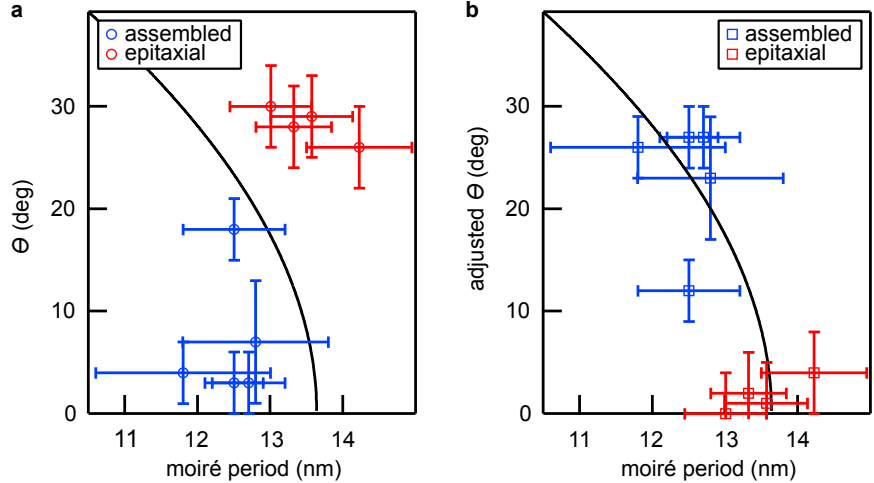

**Supplementary Figure** 6. Crystallographic orientation of stripe axes in graphene/hBN heterostructures. **a**, Blue circles indicate the measured rotational misalignment $\theta$ between stripe axes and nearest moiré lattice vectors for our five separate, nearly aligned heterostructures. Red circles indicate $\theta$ for four representative epitaxial heterostructures (of 25 measured). Black curve shows the calculated angular misalignment[3] between the moiré lattice vectors and the graphene lattice vectors, assuming that the lattice constants for hBN and graphene are $a_{hBN} = 0.25$ nm and $a_{graphene} = a_{hBN}/1.018$. Both red and blue circles should fall on this curve if the stripe axes are zigzag. **b**, Same data as in **a**, but the value plotted on the vertical axis is adjusted to be $30° - \theta$. The squares should fall on the black curve if the stripe axes are armchair.

---

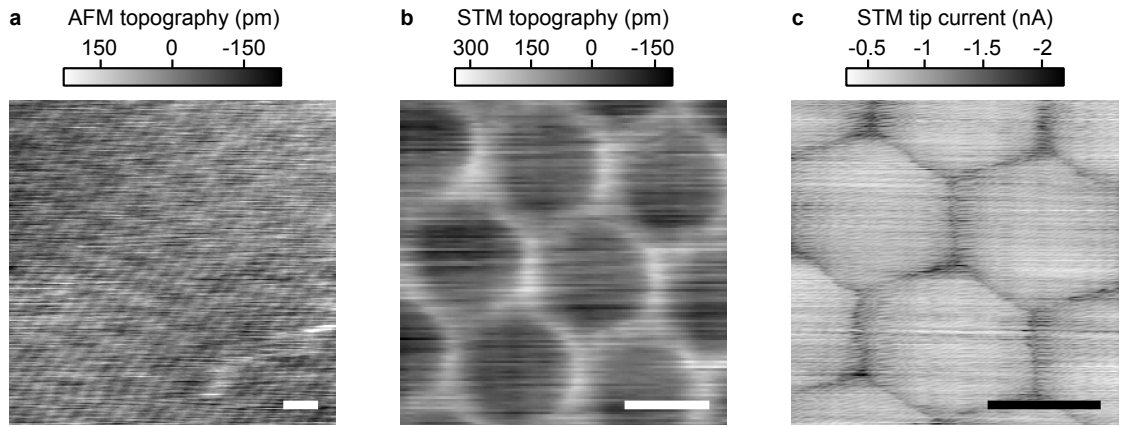

**Supplementary Figure** 7. Absence of stripes in STM. **a**, AFM topography of an epitaxial graphene/hBN heterostructure. The peak-to-trough topographic amplitudes of the moiré pattern and stripes are similar. Scale bar: 10 nm. **b**, STM topography of the same sample in constant current mode, with tip-sample bias 50 mV and current 1 nA. The moire pattern is clearly visible, with no evidence of stripes. Scale bar: 10 nm. **c**, STM topography in constant height mode, with tip-sample bias 50 mV. Again no stripes are visible, despite sharp resolution of the moiré pattern. Scale bar: 10 nm.



\end{document}